\documentclass[conference]{IEEEtran}

\usepackage{times}
\usepackage{epsfig}
\usepackage{graphicx}
\usepackage{amsmath}
\usepackage{amssymb}

\usepackage{helvet}
\usepackage{courier}
\usepackage{multirow}
\usepackage{setspace}
\usepackage{subfigure}
\usepackage{url}
\usepackage[boxed,commentsnumbered]{algorithm2e}
\usepackage{algorithmicx}
\usepackage{algpseudocode}
\usepackage{dsfont}
\usepackage{sidecap}
\usepackage{array}
\usepackage{balance}
\usepackage[pagebackref=true,breaklinks=true,letterpaper=true,colorlinks,bookmarks=false]{hyperref}

\newcolumntype{V}{>{$\vcenter\bgroup\hbox\bgroup}c<{\egroup\egroup$}}

\graphicspath{{./imgs/}}

\begin{document}

\title{Overcoming Data Scarcity of Twitter:\\ Using Tweets as Bootstrap with Application to Autism-Related Topic Content Analysis}

\author{
\begin{tabular}{ccccc}
  Adham Beykikhoshk$^\dag$ & Ognjen Arandjelovi\'c$^\ddag$ & Dinh Phung & Svetha Venkatesh\\[3pt]
\end{tabular}\\
\begin{tabular}{c}
   Centre for Pattern Recognition and Data Analytics\\
   School of Information Technology\\
   Deakin University\\
   Geelong, VIC 3216\\
   Australia\\
   \texttt{\{$^\dag$adham.beyki,$^\ddag$ognjen.arandjelovic\}@gmail.com}\\~
\end{tabular}
}

\maketitle
\thispagestyle{empty}

\begin{abstract}
Notwithstanding recent work which has demonstrated the potential of using Twitter messages for content-specific data mining and analysis, the depth of such analysis is inherently limited by the scarcity of data imposed by the 140 character tweet limit. In this paper we describe a novel approach for targeted knowledge exploration which uses tweet content analysis as a preliminary step. This step is used to bootstrap more sophisticated data collection from directly related but much richer content sources. In particular we demonstrate that valuable information can be collected by following URLs included in tweets. We automatically extract content from the corresponding web pages and treating each web page as a document linked to the original tweet show how a temporal topic model based on a hierarchical Dirichlet process can be used to track the evolution of a complex topic structure of a Twitter community. Using autism-related tweets we demonstrate that our method is capable of capturing a much more meaningful picture of information exchange than user-chosen hashtags.
\end{abstract}

\section{Introduction}\label{s:intro}
The last decade has witnessed the emergence of a number of social media platforms and the rapid rate of their proliferation, breadth of adoption, and frequency of use. The associated wealth of user-generated content, most of which is easily accessible, has created an immense opportunity for automatic data mining and knowledge discovery. It is unsurprising that this opportunity has been readily embraced both by the academic community~\cite{BaucSanjLiuChen2013,BeykAranPhunVenk+2015,Culo2010,HarsGandLazaYu+2011,HimeHan2014,LiCard2013,PaulDred2012} and the commercial sector~\cite{Hutc2012,LinRyab2013,Russ2013}. Nevertheless research in this area is still in its infancy~\cite{AsurHube2010}, with the majority of the methodologies applied to the processing of social media content to date being simple adaptations of existing algorithms developed for the analysis of `conventional' data sources. At this stage neither the full range of the associated challenges nor the potential of information found on social media are fully understood, and only when these tasks are accomplished will the development of algorithms specifically aimed at social media be able to achieve substantial progress.

\begin{figure}[thp]
  \vspace{5pt}
  \centering
  \includegraphics[width=0.48\textwidth]{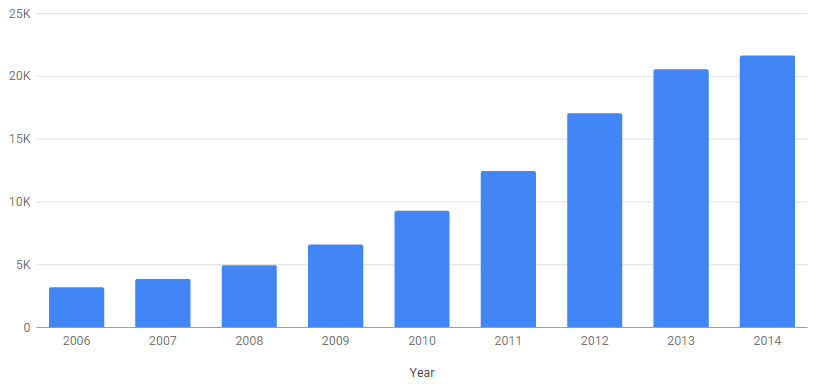}
  \caption{ Number of articles indexed by Google Scholar matching the search query \textit{twitter mining}, shown per year of publication (note that the count is \textit{per annum}, not cumulative). }
  \label{f:twitterMining}
\end{figure}

The backdrop we just summarized readily explains why Twitter in particular has attracted so much research attention, as illustrated by the plot in Fig.~\ref{f:twitterMining}. Firstly, it is immensely popular, with 288 million active users and 500 million tweets sent per day, with a high level of global adoption witnessed by the statistic that 77\% of its users are located outside the USA~\cite{Twitter}. Secondly, given that research in this field is still in its early stages, the purely textual nature of content on Twitter itself provides a sufficiently constrained environment needed to develop an understanding of the available information and the performance of different data mining and machine learning methodologies. In large part these observations motivate the focus of our work on Twitter too.

\begin{figure*}[!t]
  \centering
  \includegraphics[width=1\textwidth]{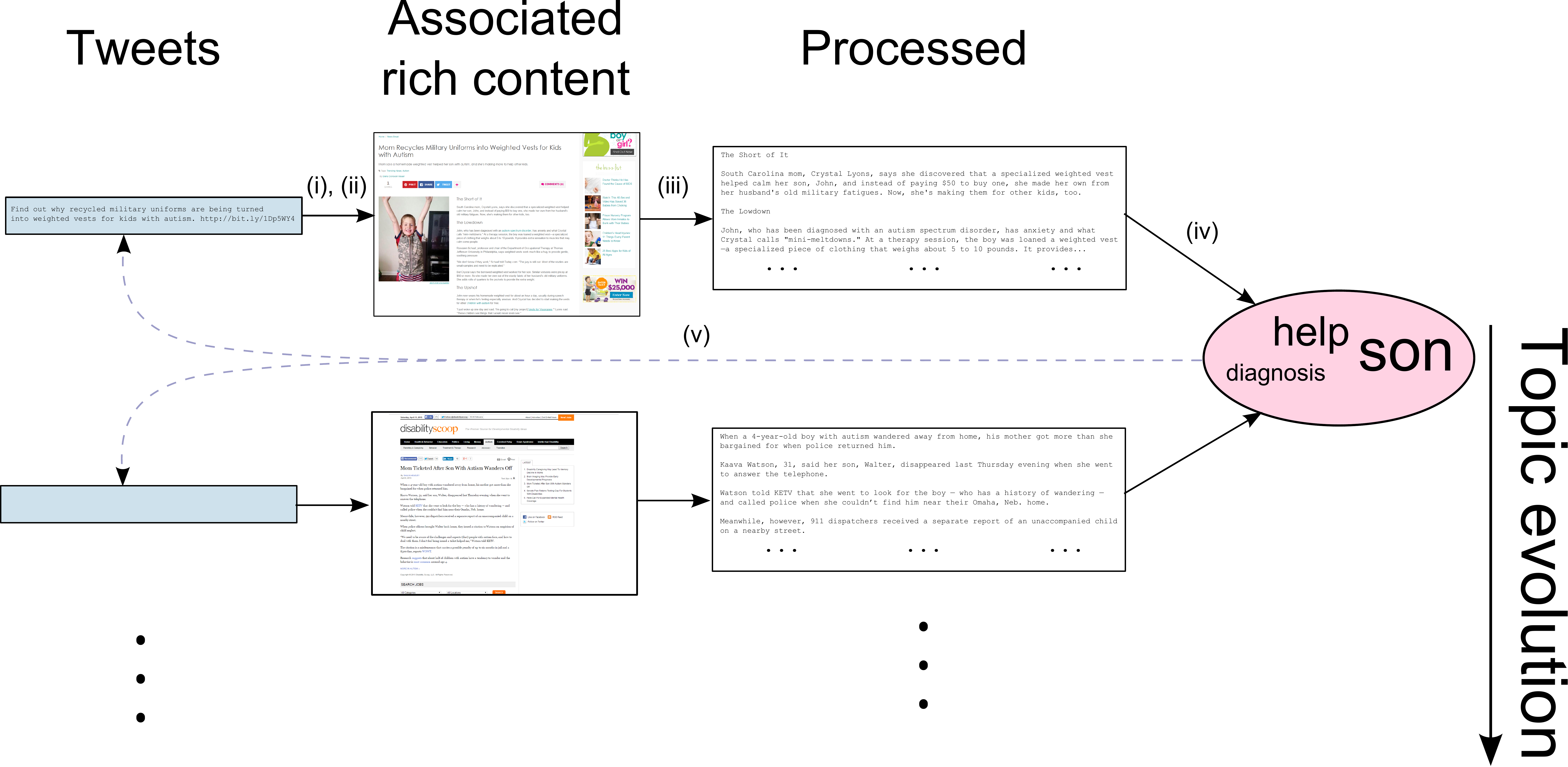}
  \caption{Conceptual overview of the proposed framework. Our analytical pipeline consists of: (i) identifying URLs in tweets, (ii) following the corresponding links, (iii) processing the target web pages to extract relevant textual information, (iv) performing topic analysis on these documents, and finally (v) linking the extracted topic back with the original tweets.
  }
  \label{f:mainIdea}
\end{figure*}

The specific interest of the present work lies in the use of Twitter content for the monitoring and analysis of attitudes, behaviours, and conversations pertaining to health care and the autism spectrum disorder (ASD)~\cite{DSM5} in particular. Our interest in this condition is motivated by two key factors. The first of these concerns the associated human and economic cost~\cite{JacoMuliGree1998}. In brief, ASD is a life-long neurodevelopmental disorder~\cite{LevyMandSchu2009} characterized by severe impairments in social interaction, communication, and in some cases cognitive abilities~\cite{Mile2011}. ASD typically begins already in infancy or at the very latest by the age of three when it is usually detected by an abnormal lack of social reciprocity. Current evidence suggests that a stable proportion of approximately 0.5--0.6\% of the population is afflicted by ASD although the actual diagnosis rate is on the increase due to the broadening diagnostic criteria~\cite{BaxtBrugErskSche+2015}.

The second important factor motivating our research stems from the potential of Twitter as a pervasive interactive platform to reach ASD communities which are notoriously difficult to engage by traditional means. This includes both those segments of the ASD community which are isolated socially, for example due to stigma often attached to mental disorders~\cite{Gray1993}, as well as those which are isolated geographically, for example in remote rural communities. From the public policy point of view, understanding the practices and beliefs of parents and carers of ASD-affected individuals is crucial, yet often lacking~\cite{HarrRoseGarnPatr2006}. The penetration of beliefs, and behavioural and educational interventions which are questionable~\cite{WarrMcPhSathFoss2011} and poorly supported by evidence (e.g.\ gluten-free and casein-free diets, and cognitive behavioural therapy~\cite{DaniWood2013}), and sometimes outright in conflict with science~\cite{TremBalaRoss2005} is particularly worrying~\cite{WarrMcPhSathFoss2011}.

Thus the key premise behind our research efforts is that the rapid rise in the adoption of social media as a platform for the expression and exchange of ideas, which facilitates the emergence of special interest communities, can be used to study and monitor the beliefs and practices of the population affected by ASD. Considering the challenge of reaching and engaging with this specific target population, success in this domain of data mining would be a powerful utilization of social networks for public good.


\subsection{Related previous work}
Most of the previous work on data mining Twitter has not been automatic, that is, analysis was performed `manually' by humans within the proof-of-concept paradigm. Considering the usage statistics summarized in Sec.~\ref{s:intro}, it is evident that non-machine-based approaches are not feasible in practice -- the process is extremely laborious which severely limits how much data can be processed. A major challenge in automating the analysis lies in the likely semantic gap needed to interpret Twitter messages. As we shall demonstrate in the next section, the severe limit on the character length of tweets appears to exaggerate this gap yet further, requiring an even more sophisticated understanding of language, its idioms, sociolects etc. The following ASD-related tweets, representative of the corpus we collected (see Sec.~\ref{s:eval}), readily illustrate this:\\~\\
\textbf{Example tweet 1:}~
\begin{quote}\it
  Looks like we will have more \#autism research happening for children in \#EarlyIntervention next year! :-) \#VisualSupports \#MobileTech
\end{quote}
\textbf{Example tweet 2:}~
\begin{quote}\it
  Authors who see autism as ``tremendously burdening'' elicit dire views of autism from parents http://j.mp/1FyqXwF  ``Ethical approval: none''
\end{quote}
\textbf{Example tweet 3:}~
\begin{quote}\it
  101 autism: Genetic analysis of individuals with autism finds gene deletions - Using powerful genetic sequencing. http://is.gd/UhprQK
\end{quote}
\textbf{Example tweet 4:}~
\begin{quote}\it
  \#Apple \#Censorship \& Dr. Brian Hooker Interview exposing CDC Cover-up of the Vaccine \& Autism Link on .@rediceradio  http://youtu.be/19uvPtg6SPI
\end{quote}

Although of limited practical use in the long term, manual approaches to Twitter data analysis and mining have served an important role in illustrating the potential of this platform. Indeed the range of applications in which the usefulness of Twitter data has been demonstrated is remarkable. For example, the inference of message sentiment from emoticons was described by Jiang \textit{et al.}~\cite{JianYuZhouLiu+2011}, and Bifet and Frank~\cite{BifeFran2010}, while Agarwal \textit{et al.}~\cite{AgarXieVovsPass2011} employed linguistics-based classifiers to a similar end. Sentiment analysis in tweets was further used by Bollen \textit{et al.} \cite{BollMaoPepe2011} in the prediction of socio-political, cultural, and economic events. Sociometric applications were also explored by Mitchell \textit{et al.} \cite{MitcFranHarrDodd+2013}. The potential of Twitter in the understanding and management of emergency situations, such as earthquakes, attracted significant research attention too \cite{VermViewCorvPale+2011}, with successes reported by Robinson \textit{et al.}~\cite{RobiPoweCame2013} and Sakaki \textit{et al.}~\cite{SakaOkazMats2010} amongst others.

The same themes drawing from the use of geographical information and the real-time nature of Twitter can be observed in the existing work on the use of the platform in the realm of health care \cite{ChewEyse2010}. For example Paul and Dredze \cite{PaulDred2011} tracked over time the spread of illnesses, changes in behavioural risk factors, and symptoms and patterns of medication use. Patterns of antibiotic use were analysed by Scanfeld \textit{et al.} \cite{ScanScanLars2010} while Jashnisky \textit{et al.}~\cite{JashBurtHansWest+2014} studied the possibility of predicting suicide rates.


Notwithstanding the increasingly recognized need for more effective information exchange with the affected communities, the use of social media in the understanding and management of ASD has received little research attention. A manual (i.e.\ human effort-based, rather than computerized) analysis of ASD-affected individuals' writing patterns was reported by Newton \textit{et al.}~\cite{NewtKramMcIn2009} while our previous work~\cite{BeykAranPhunVenk+2014} described the only automatic system in this domain.

\section{Proposed method}
In this section we lay out the key contributions of the present work. We start with an overview of the proposed methodology and then proceed with a detailed description of each of its components in turn.

\subsection{Data mining framework -- key ideas}\label{ss:keyIdeas}
In this paper our aim is to extract the topic content of a specific category of tweets and track its changes over time. Although our focus is on ASD-related tweets, which we showed can be readily detected automatically~\cite{BeykAranPhunVenk+2014}, the nature of the proposed framework is entirely general and could in principle be applied to any other category of tweets. Here `topic' (and any derivations thereof) is used as a technical concept from the field of text analysis, that is, it refers to a probability distribution over a collection of words/terms (the `dictionary')~\cite{BleiLaff2006}. Intuitively, this can be interpreted as a formalization of the colloquial meaning of the word `topic'~\cite{ChanGerrWangBoyd+2009}.

Probably the most obvious approach to tweet topic modelling involves the use of Twitter's built-in capability for tagging tweets by their authors, using so-called `hashtags'. These can be recognized by the leading special character `\#' (the hash sign) and are understood to be meta-data labels. We found that approximately 25\% of ASD-related tweets contain hashtags. However recent research has shown that a small number of keywords is inadequate in capturing the essence of the content of the associated tweet~\cite{BleiLaff2006}.

\begin{figure}[htb]
  \centering
  \subfigure[Daily tweeting rate]{\includegraphics[width=.48\textwidth]{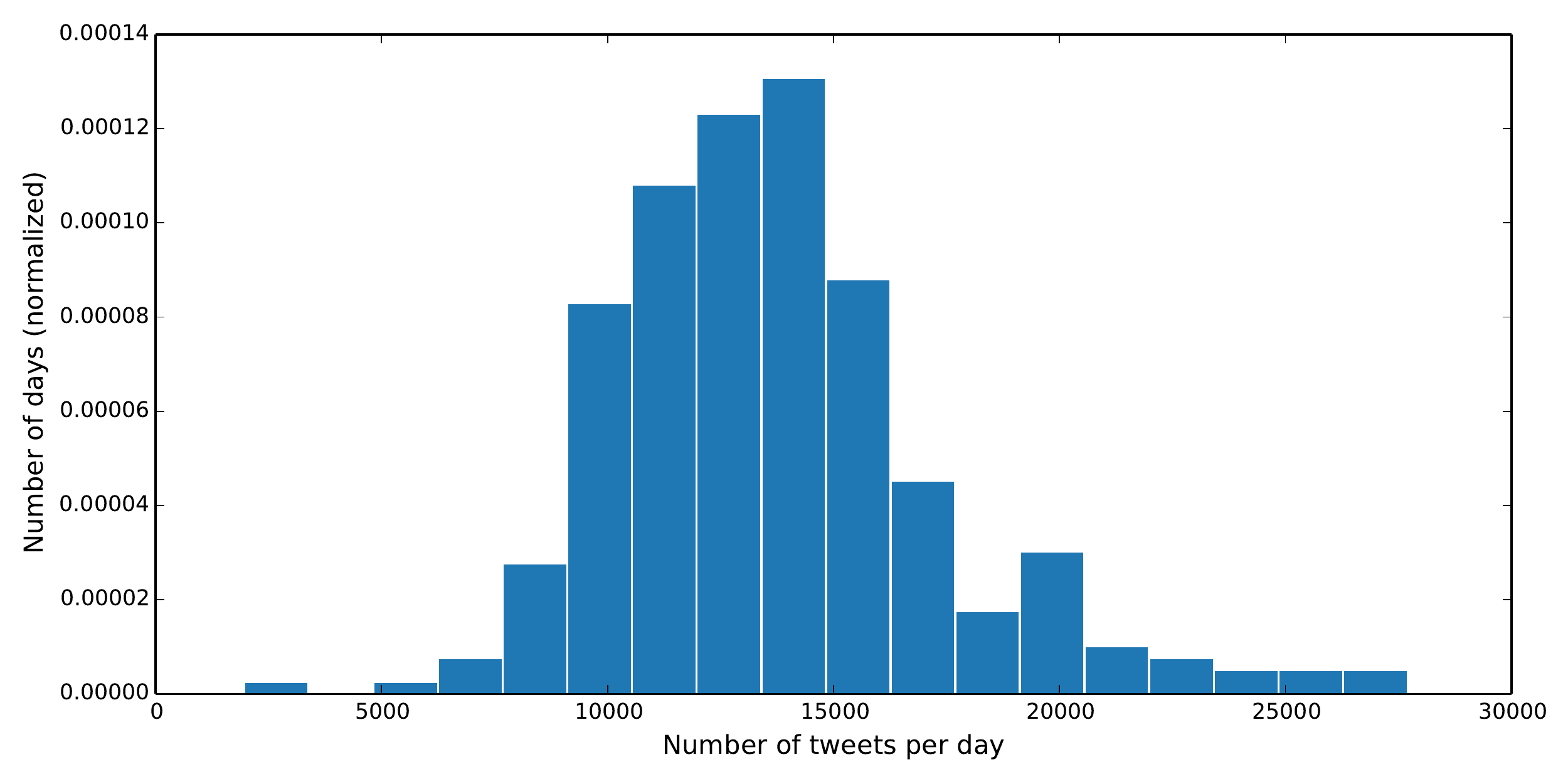}}
  \subfigure[Daily in-tweet URL rate]{\includegraphics[width=.48\textwidth]{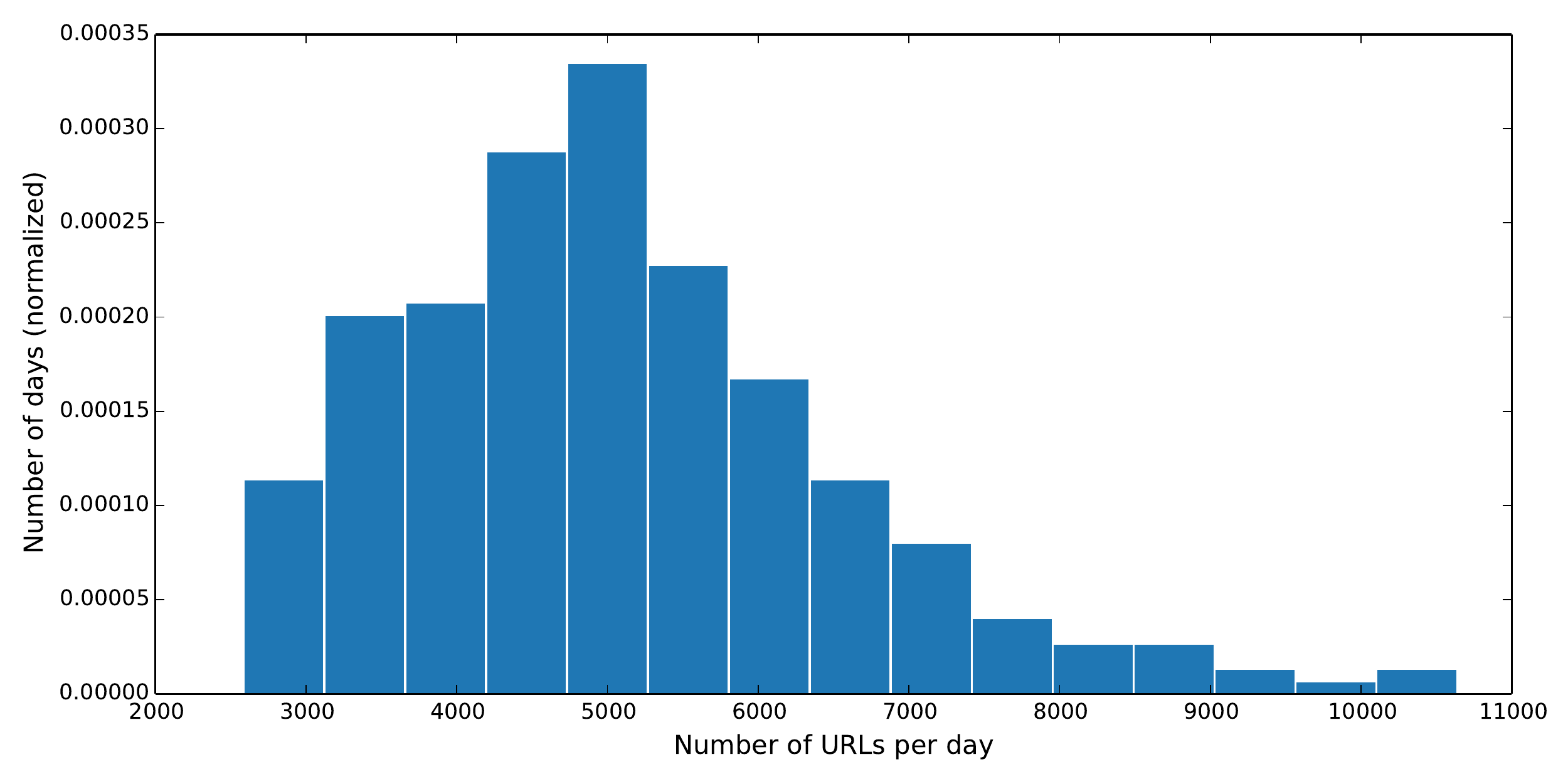}}
  \caption{The statistics of relevant data availability rate, estimated from our data set (please see Sec.~\ref{s:eval} for additional detail). Shown are histograms corresponding to (a) the daily number of tweets identified as ASD-related~\cite{BeykAranPhunVenk+2014}, and (b) the daily count of URLs harvested from these tweets. }
  \label{f:stats}
\end{figure}

The idea that the main textual information within tweets can be analysed for its topic content is motivated by the findings reported in our previous work~\cite{BeykAranPhunVenk+2014}. In particular we showed that even after terms which are \textit{ipso facto} ASD-related are removed, there remains sufficient information (both structural and semantic) which is characteristic to ASD-related tweets. However our attempts at applying a variety of different topic models failed to produce meaningful results. Specifically we found that the number of terms contained in individual tweets (the average term count per tweet in our corpus is 13) is too small to give rise to topics of sufficient complexity. Considering that this bottleneck is inherent to Twitter rather than a limitation of our methodology, at first sight it appears that topic content extraction from tweets is impossible, to say nothing of any temporal analysis thereof. However, in this work we show that this barrier can be overcome through an ingenious trick. In particular we observed that a large number of tweets contain URLs i.e.\ links to external content, as illustrated in Fig.~\ref{f:stats}. Moreover these tweets are exactly those that we are most interested in, that is, those which aim to contribute the greatest amount of information.

Hence our idea is to use tweets not as endpoint data mining sources themselves but rather as data mining intermediaries used to discover much richer associated content. Some previous work has already recognized the potential of such external content e.g.\ \cite{SankSameTeitLieb+2009,AbelGaoHoubTao2011}. However, unlike Sankaranarayanan \textit{et al.}~\cite{SankSameTeitLieb+2009} who use linked content for clustering and the identification of popular news topics, or Abel \textit{et al.}~\cite{AbelGaoHoubTao2011} who focus on building news-related user models, we demonstrate that content harvested from linked URLs can be used to build a complex, temporal and cross-modal graph which facilitates temporal topic analysis which can be mapped back to Twitter posts (an interesting comparison can be made with machine learning methods in other fields which exploited such parallel structures with success e.g.\ for inferring pose-wise mappings across face manifolds~\cite{AranCipo2013,Aran2012}). In brief, our analytical pipeline consists of: (i) identifying URLs in tweets, (ii) following the corresponding links, (iii) processing the target web pages to extract relevant textual information, (iv) performing topic analysis on these documents, and finally (v) linking the extracted topic back with the original tweets. This is illustrated conceptually in Fig.~\ref{f:mainIdea}. Our automatic extraction of relevant textual information from web pages is described in the next section. A method for tracking complex temporal changes of the topic structure of a longitudinal document corpus, based on a graph constructed over topics extracted using a hierarchical Dirichlet process-based model for different epochs is explained in Sec.~\ref{ss:evolution}.

\subsection{Automatic data enrichment using URLs}
As indicated in the previous section, the key idea of the proposed method for Twitter topic analysis and tracking is to map tweets which contain URLs to the space of the corresponding web pages. This process effectively establishes correspondence between a tweet and the web page linked through a URL in the tweet body. Observe that in principle a single tweet can be put in correspondence with multiple web pages as multiple URLs may be contained within the tweet, though in practice this is seldom the case -- we found that over 96\% of URL-containing tweets include a single URL, and fewer than 0.1\% more than two URLs. Equally, a single web page can be linked to from different tweets; this is a far more frequent occurrence. Whatever the exact network of correspondences may be, topics assigned to the content of a specific web page are also associated with all tweets which contain a URL link to this web page. As per our previous observations, this means that a tweet with multiple URLs `inherits' topics from multiple web pages, and equally, a web page lends its topics to all tweets which link to it.

We retrieve tweets using Twitter's stream and search APIs (please see Sec.~\ref{s:eval} for further detail). This allows us to fetch relevant tweet information, such as the text of the message, posting time and date, geo-tags, and the author's user ID, as well as the list of any URLs linked to from the tweet in the JavaScript Object Notation (JSON) format. Therefore no sophisticated processing of the tweet is needed at this stage -- the URLs of interest are retrieved explicitly. These are then used to fetch the corresponding web pages. We process only those URLs which link to HTML pages. For example by examining the `Content-Type' response from the HTTP server we immediately reject links to videos and other non-text-based content. Pages which satisfy our validity criteria are then processed to extract their main text content. This is achieved by analysing the corresponding HTML descriptions and removing chunks of it which by their HTML tags can be identified as not being of interest. For example we remove text enclosed within tags such as \texttt{<COMMENT>}, \texttt{<META>}, \texttt{<HEADER>}, \texttt{<MENU>}, \texttt{<RSS>}, \texttt{<SPONSOR>}, and a series of others. Formatting tags such as \texttt{<FONT>}, \texttt{<DIV>}, \texttt{<P>}, and similar, are merely removed. We actively retain text within tags such as \texttt{<MAIN>}, \texttt{<ARTICLE>}, and \texttt{<BLOG>}, amongst others. This processes allows us to extract relevant content reliably without any manual input, with few errors. We found that rare errors do not affect the ultimate result of our method due to their randomness -- there is no reason why any particular erroneous term (word) would consistently be included in the extracted text, which means that any \emph{specific} error ends up being treated as noise in our bag-of-words model used as input to the main topic modelling and tracking algorithm described next.

\subsection{Modelling topic evolution over time}\label{ss:evolution}
Twitter is by its very nature highly dynamic: new content is rapidly created, transient periods of high popularity of specific ideas are followed by their rapid decline and the emergence of follow-up etc. Therefore it is imperative for the analysis of the topic structure of a tweet corpus to have a temporal component, which additionally has to be sufficiently robust and flexible to capture the range of complex interactions and causal cascades that the corresponding exchanges exhibit.

Our approach begins by discretizing time into epochs thereby dividing the tweet corpus into a series of chronological but overlapping sub-corpora. The topic structure of each sub-corpus associated with a particular epoch is then extracted separately using an HDP-based model. In other words each epoch is considered to span a sufficiently short time period that the associated sub-corpus of tweets can be treated as a static collection of documents. To ensure the validity of this assumption it is crucial that the time span of an epoch is chosen based on the speed of changes which the model is supposed to track (to reflect the highly dynamic nature of Twitter in this work we used three day epochs, with an overlap of successive epochs of two days). To associate topics across different epochs we create a similarity graph, connecting topics (as nodes) across successive epochs, and infer complex dynamics based on the strengths of these connections. This approach, first proposed in~\cite{BeykPhunAranVenk2015}, is explained next.

\subsubsection{Measuring inter-topic similarity}
The key idea behind our approach stems from the observation that while topics may change significantly over time, by their very nature the change between successive epochs is limited. Therefore we infer the continuity of a topic in one epoch by relating it to all topics in the immediately subsequent epoch which are sufficiently similar to it under some similarity measure. This can be seen to lead naturally to a similarity graph representation whose nodes correspond to topics and whose edges link those topics in two epochs which are related. Formally, the weight of the directed edge that links $\phi_{j,t}$ , the $j$-th topic in epoch $t$, and $\phi_{k,t+1}$ is set equal to $\rho\left(\phi_{j,t},\phi_{k,t+1}\right)$ where $\rho$ is an appropriate similarity measure. In this work we adopt the use of the Jaccard similarity between the probability distributions described by $\phi_{j,t}$ and $\phi_{k,t+1}$ i.e.\ $\rho\left(\phi_{j,t},\phi_{k,t+1}\right) \equiv d_\text{Jaccard}\left(\phi_{j,t},\phi_{k,t+1}\right)$.

\begin{figure}
  \centering
  \includegraphics[width=0.36\textwidth]{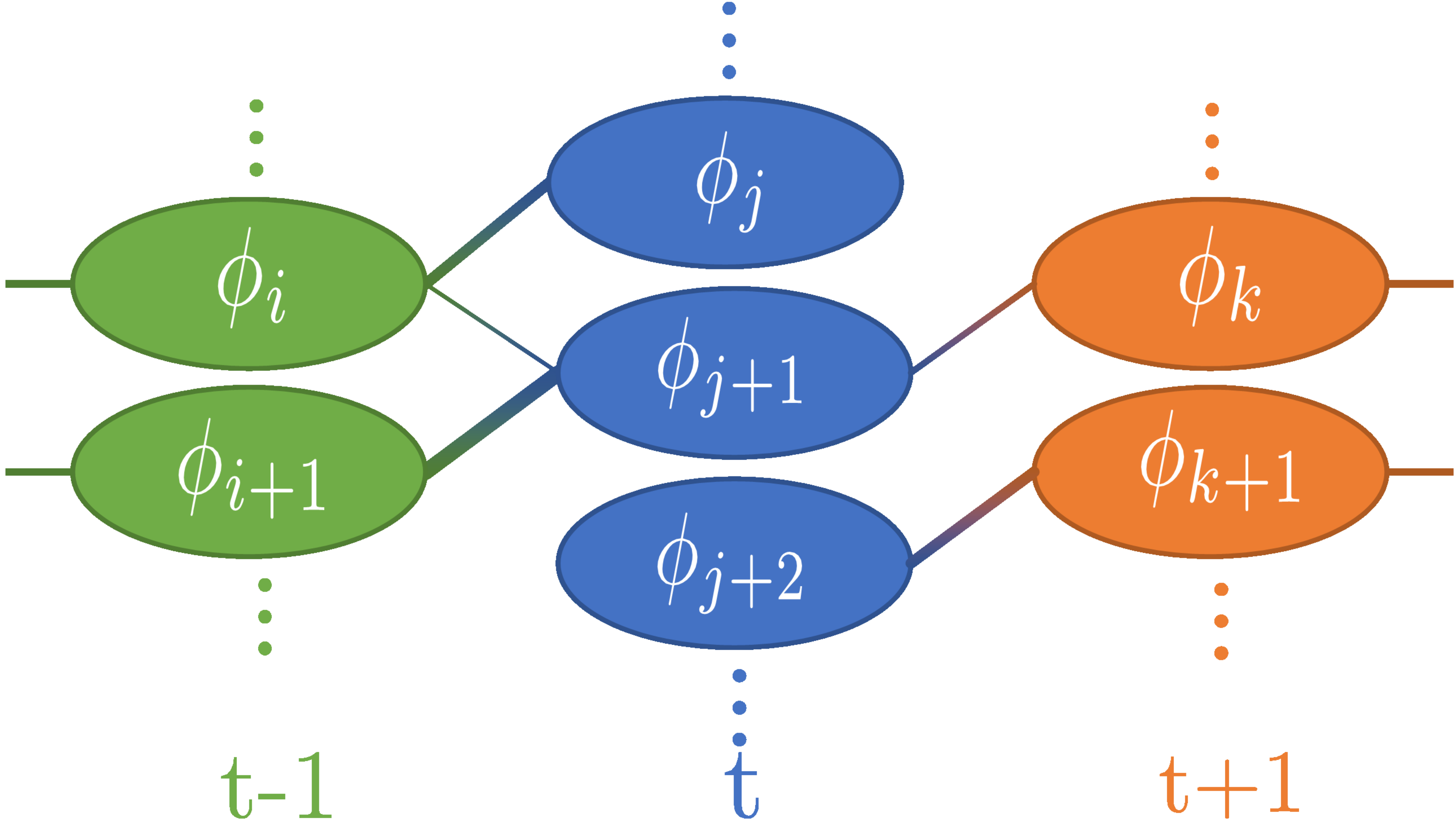}
  \vspace{5pt}
  \caption{A conceptual illustration of a similarity graph that models topic dynamics over time. A node corresponds to a topic in a specific epoch; the weight of an edge connecting two nodes is equal to the similarity of the corresponding topics.}
  \label{f:evolution}
\end{figure}

A conceptual illustration of a similarity graph is shown in Fig.~\ref{f:evolution}. It shows three consecutive time epochs $t-1,t,$ and $t+1$ and a selection of topics in these epochs. Graph edge weight i.e.\ inter-topic similarity is encoded by varying the thickness of the corresponding line connecting two nodes -- a thicker line signifies more similar topics. We use a threshold to eliminate weak edges automatically, retaining only the edges which correspond to sufficiently similar topics in adjacent epochs. It can be seen that this readily allows us to detect the disappearance of a particular topic, the emergence of new topics, as well as the splitting or merging of different topics. In summary:\\[-10pt]
\begin{list}{}{\leftmargin=20pt}
  \item[\bf Emergence] If a node does not have any edges incident to it, the corresponding topic is taken as having emerged in the associated epoch (e.g.\ $\phi_{j+2}$ at time $t$ in Fig.~\ref{f:evolution}).\\[-0pt]
  \item[\bf Disappearance] If no edges originate from a node, the corresponding topic is taken to vanish in the associated epoch (e.g. $\phi_{j}$ at time $t$ in Fig.~\ref{f:evolution}).\\[-0pt]
  \item[\bf Splitting] If more than a single edge originates from a node, the corresponding topic is understood as being split into multiple topics in the next epoch (e.g.\ $\phi_{i}$ is split into $\phi_{j}$ and $\phi_{j+1}$ in Fig.~\ref{f:evolution}).\\[0pt]
  \item[\bf Merging] If more than a single edge is incident to a node, the topics of the nodes from which the edges originate are understood as having merged together to form a new topic (e.g.\ $\phi_{i}$ and $\phi_{i+1}$ merge to form $\phi_{j+1}$ in Fig.~\ref{f:evolution}).
\end{list}

\subsubsection{Hierarchical Dirichlet process mixture models}\label{sss:hdp}
The Dirichlet process is a widely used prior for mixture modelling which readily allows a document collection to accommodate an arbitrary number of topics. It is the building block of many Bayesian nonparametric methods~\cite{Ferg1973}. Succinctly put, a Dirichlet process $\text{DP}\left(\gamma,H\right)$ is defined as a distribution of a random probability measure $G$ over a measure space $\left(\Theta,\mathcal{B},\mu\right)$, such that for any finite measurable partition $\left(A_{1},A_{2},\ldots,A_{r}\right)$ of $\Theta$ the random vector $\left[G\left(A_{1}\right),\ldots,G\left(A_{r}\right)\right]^T$
is a Dirichlet distribution with parameters $\left[\gamma H\left(A_{1}\right),\ldots,\gamma H\left(A_{r}\right)\right]^T$.

Owing to the discrete nature and infinite dimensionality of its draws, the DP is a highly useful prior for Bayesian mixture models. By associating different mixture components with atoms $\phi_{k}$ of the so-called stick-breaking process~\cite{IshwJame2001}, and assuming $x_{i}|\phi_{k}\overset{iid}{\sim}F\left(x_{i}|\phi_{k}\right)$ where $F\left(.\right)$ is the likelihood kernel of the mixing components, the Dirichlet process mixture model (DPM) can be formulated. The DPM is suitable for nonparametric clustering of exchangeable data in a single group e.g.\ words in a document where the DPM models the underlying structure of the document with potentially an infinite number of topics. However, many real-world problems are more appropriately modelled as comprising multiple groups of exchangeable data (e.g.\ a collection of documents). In such cases it is usually desirable to model the observations of different groups jointly, allowing them to share their generative clusters. This idea is known as the sharing statistical strength and is achieved using a hierarchical structure.

Amongst different ways of linking group-level DPMs, HDP~\cite{TehJordBealBlei2004} offers an interesting solution whereby base measures of document-level DPs are drawn from another DP. In this way the atoms of the corpus-level DP (i.e.\ topics in our case) are shared across the corpus. Formally, if $\mathbf{x}=\left\{ \mathbf{x}_{1},\ldots,\mathbf{x}_{J}\right\} $ is a document collection where $\mathbf{x}_{j}=\left\{ x_{j1},\ldots,x_{jN_{j}}\right\} $ is the $j$-th document comprising $N_{j}$ words, each document is modelled with a DPM $G_{j}|\alpha_{0},G_{0}\stackrel{iid}{\sim}\text{DP}\left(\alpha_{0},G_{0}\right)$ where its DP prior is further endowed by another DP $G_{0}|\gamma,H\sim\text{DP}\left(\gamma,H\right)$. Since the base measure of $G_{j}$ is drawn from $G_{0}$, it takes the same support as $G_{0}$. Also the parameters of the group-level mixture components, $\theta_{ji}$, share their values with the corpus-level DP support on $\left\{ \phi_{1},\phi_{2},\ldots\right\} $. The posterior for $\theta_{ji}$ follows a Chinese restaurant franchise process which can be used to develop inference algorithms based on Gibbs sampling~\cite{TehJordBealBlei2006} which we adopt in the present work.

\section{Evaluation and results}\label{s:eval}
In this section we turn our attention to the empirical evaluation of the proposed framework. We first succinctly describe our evaluation data set, and then report and discuss the obtained results.

\subsection{Evaluation data}\label{ss:data}
Twitter's Terms of Service explicitly prohibit the sharing or redistribution of tweets, including for research purposes. Consequently, as there was no public benchmark that we could adopt we used a large data set collected by ourselves, acquired as described in~\cite{BeykAranPhunVenk+2014}. For full detail the reader is referred to the original publication. Here we note the sole difference in the tweet collections used in the present work and that in~\cite{BeykAranPhunVenk+2014}: their sizes.

Recall that we collected tweets using Twitter's stream and search APIs by periodically retrieving tweets which contain any of the four keywords ``autism'', ``adhd'', ``asperger'', and ``aspie'' (or any of their derivatives obtained by suffixation). Following the publication of our original work we continued this process and now have at our disposal a data set over an order of magnitude greater than that initially described in~\cite{BeykAranPhunVenk+2014}. Our corpus now includes 5,650,989 ASD-related tweets collected in the period starting on 26 August 2013 and ending on 1 Oct 2014 (i.e.\ more than 13 consecutive months).

\subsubsection{Algorithm parameters}

The analytics framework we described in the preceding sections requires the values of a number of free parameters to be set. For the sake of completeness and full reproducibility of the experiments reported in this paper, here we summarize how this was performed and what values were used. Note that this process needs to be performed only once -- the framework is fully automatic thereafter. In addition, in the discussion of possible directions for future work in Sec.~\ref{s:summary} we also outline the ideas we are currently working on which would allow the optimal values of the free parameters to be learnt directly from data. This would make the entire process automatic and remove the need for human input even to the minimal extent described here.

The first of the parameters, introduced in Sec.~\ref{ss:keyIdeas}, is the size of the term dictionary used to represent documents as fixed-size bags of words/terms. It should be noted that in principle the entire set of terms extracted from the document corpus could be used without any negative effects on the quality of the extracted topic information, as the importance of each term is inferred automatically by our HDP-based model. The sole reason for choosing a subset of terms as a dictionary lies in the reduced computational overhead (both time-wise and storage-wise). Following the usual procedure adopted in the literature (for detailed discussion see e.g.~\cite{BeykAranPhunVenk+2015}) we construct the dictionary using the most frequent terms which account for approximately 90\% of the text of the document corpus~\cite{ClauShalNewm2009}. In our case this results in 6560 terms. It is insightful to observe the much greater size of this dictionary than the 1500-term one constructed in exactly the same manner in \cite{BeykAranPhunVenk+2015} but using main tweet text only. This reflects the more expressive nature of content linked to from tweets compared to the content within tweets themselves.

The remaining free parameters concern the proposed algorithm for temporal tracking of topic structure changes of the document corpus. In particular, to reflect the highly dynamic nature of Twitter we chose short epochs, each spanning three days, with successive epochs exhibiting an overlap of two days. Furthermore, for the pruning of inter-epoch similarity graphs we used the Jaccard similarity threshold of $\tau_\text{prune}=0.5$ which was determined empirically using a small training corpus and qualitative judgement (see Sec.~\ref{s:summary} for an indication of how this threshold can be learnt directly from data).

\subsection{Results and discussion}\label{ss:res}
We started our analysis of the results obtained with the proposed method by examining the topic structures discovered with different epochs. As in all related work, given that the technical term `topic' is understood to be a formalization and a proxy for the more abstract notion of a `topic' in everyday speech, this analysis is inherently subjective in nature as it is not possible to define or indeed obtain an objective ground-truth.

\begin{figure}[htb]
  \centering
  \subfigure[Example topic 1]{\fbox{\includegraphics[width=0.22\textwidth]{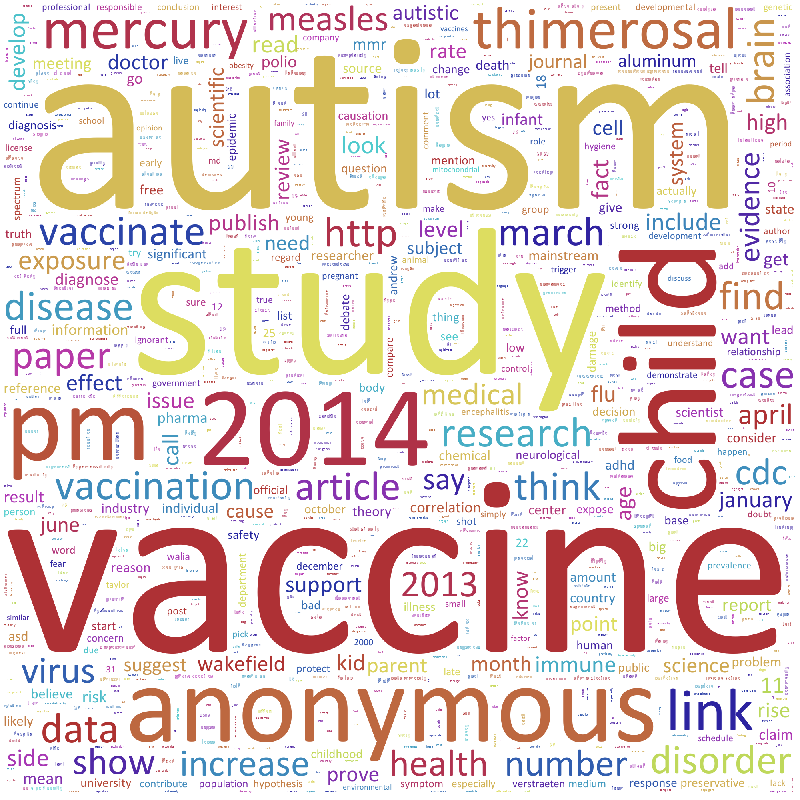}}}~~~
  \subfigure[Example topic 2]{\fbox{\includegraphics[width=0.22\textwidth]{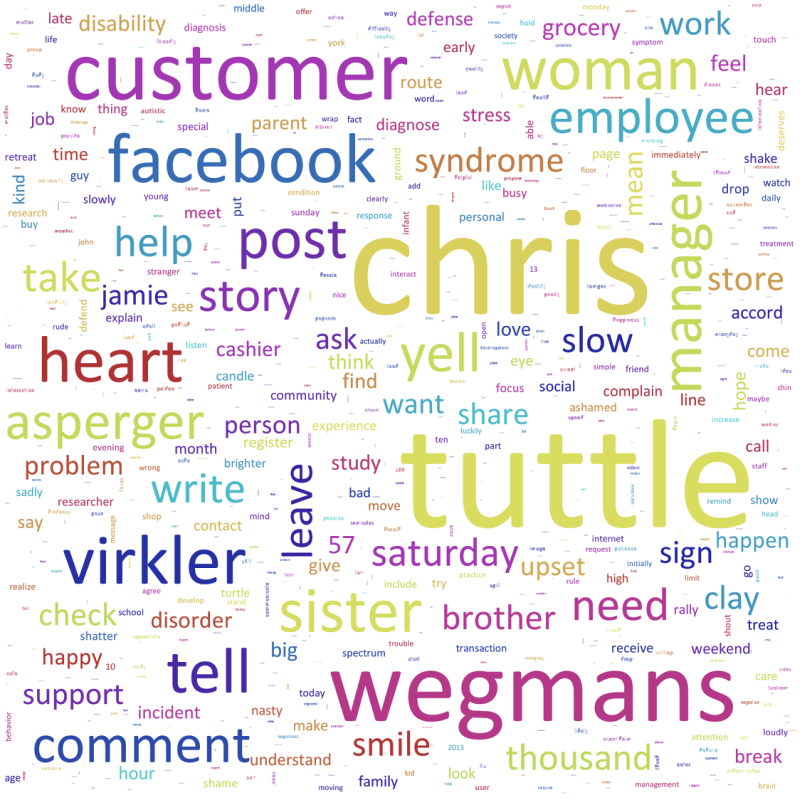}}}
  \subfigure[Example topic 3]{\fbox{\includegraphics[width=0.22\textwidth]{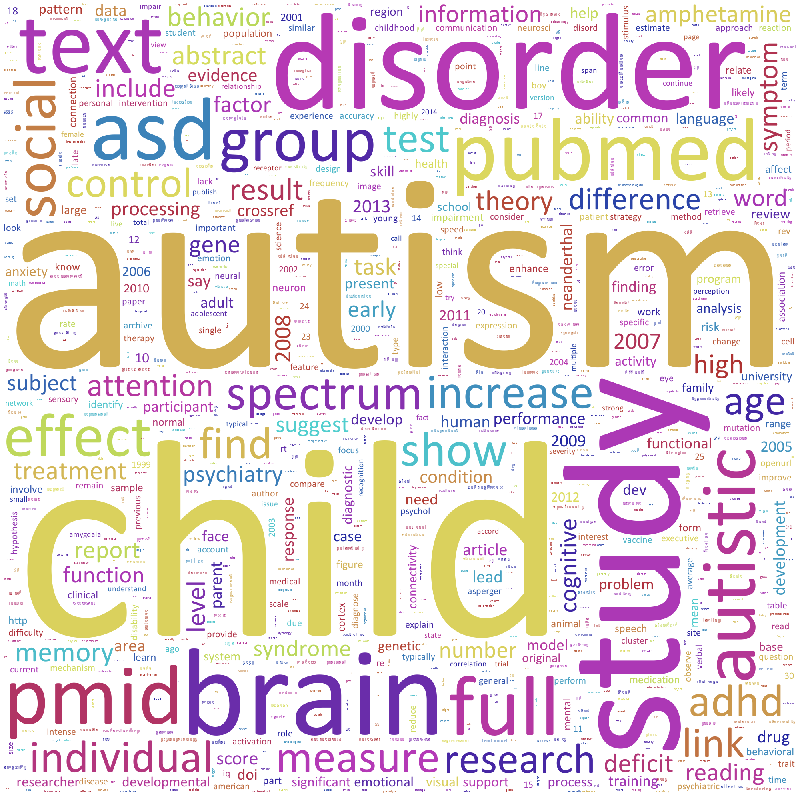}}}~~~
  \subfigure[Example topic 4]{\fbox{\includegraphics[width=0.22\textwidth]{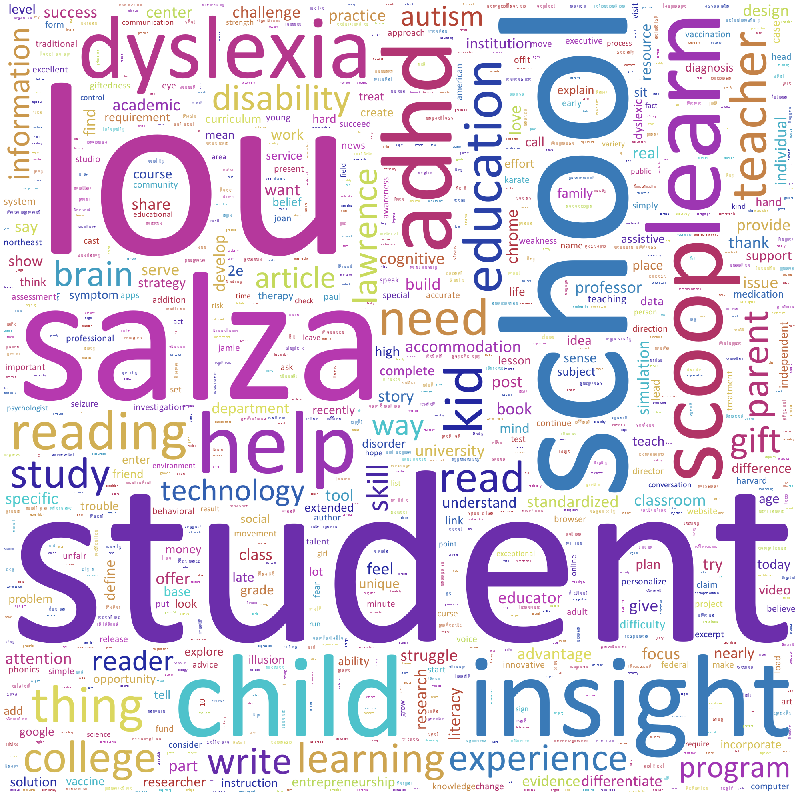}}}
  \vspace{0pt}
  \caption{ An illustration of typical topics discovered by the proposed method in a single epoch of our ASD tweet dataset, shown as size-coded word-clouds -- a larger font indicates a proportionally more probable term (different colours are merely used for easier visualization and encode no information pertaining to the corresponding terms themselves). Topic~1 can be readily related to the persistent myth of a link between childhood vaccination and autism development as well as related issues concerning the effects of thimerosal and mercury~\cite{HviiStelWohlMelb2003}, Topic~2 to the incident involving Chris Tuttle, a Wegmans employee
  with Asperger's syndrome who attracted media attention in November 2013, Topic~3 to medical literature on autism, children, and brain development, and Topic~4 to the schooling of children with learning disabilities. }
  \label{f:topics}
\end{figure}

A representative selection of popular topics discovered within an epoch is shown in Fig.~\ref{f:topics}. This figure shows four representative topics by the proposed method in a single epoch of our ASD tweet dataset. The topics are visualized using the standard world-cloud representation whereby term frequency is proportional to the corresponding font size (note that different colours are merely used for the sake of easier visualization of a large number of terms of different frequencies in limited space; colour encodes no pertinent information about different terms). There are two key observations that are important to make here. The first of these is that the extracted topics are readily interpreted as being meaningful in the context of what is known about autism. For example, Topic~1 can be seen to relate to the persisting myth of a link between childhood vaccination and autism development~\cite{Gros2009}. Related issues which concern the effects of thimerosal (a vaccine preservative) and mercury which feature in this topic are also related to this myth~\cite{HviiStelWohlMelb2003}. Topic~2 concerns Chris Tuttle, a man with Asperger's syndrome working in Wegmans who attracted much media attention in November 2013 after being yelled at by a customer. Topic~3 captures discussions of research on the relationship between autism and various health conditions and reflects the increased willingness of the general public, particularly those affected by rare and debilitating conditions, to examine the scientific literature. Lastly Topic~4 pertains to the schooling of children with learning disabilities which is one of the most common concerns of many parents with autistic children (Lou Salza, whose name features significantly in this topic, is the outspoken headmaster of a school for children with learning disabilities). The second important observation, one which strongly supports the main premise of the present work and demonstrates the usefulness of the proposed framework, is that the topics feature more than a couple of dominant (by frequency) terms. This means that they cannot be readily captured using a small set of keywords such as hashtags or words extracted directly from tweets.

\begin{figure}[htb]
  \centering
  \subfigure[Example topic progression 1 (linear)]{\includegraphics[width=0.48\textwidth]{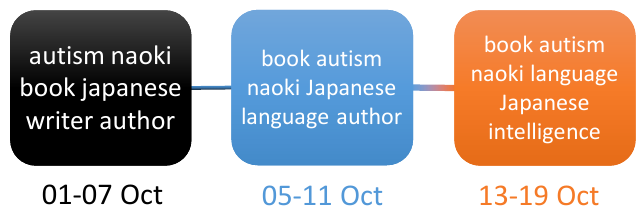}\label{f:progress_a}}
  \subfigure[Example topic progression 2 (linear)]{\includegraphics[width=0.48\textwidth]{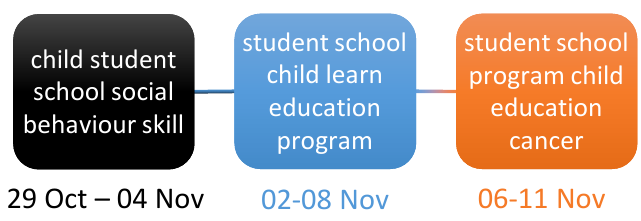}\label{f:progress_b}}
  \subfigure[Example topic progression 3 (merging)]{\includegraphics[width=0.48\textwidth]{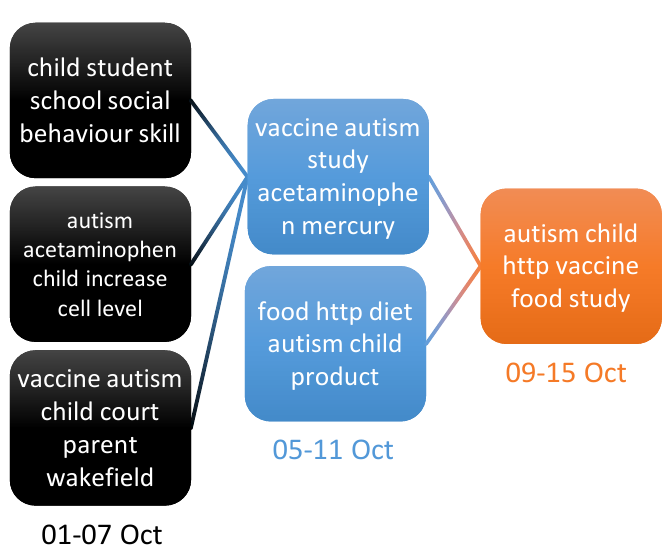}\label{f:progress_c}}
  \caption{Examples of automatically discovered Twitter topic temporal changes: (a,b) linear evolution of topic content, and (c) topic creation through the merging of topics from the previous epoch. Also see Fig.~\ref{f:progressStats} for related quantitative corroboration of our findings. }
  \label{f:progress}
\end{figure}

\begin{figure}[!t]
  \centering
  \includegraphics[width=0.48\textwidth]{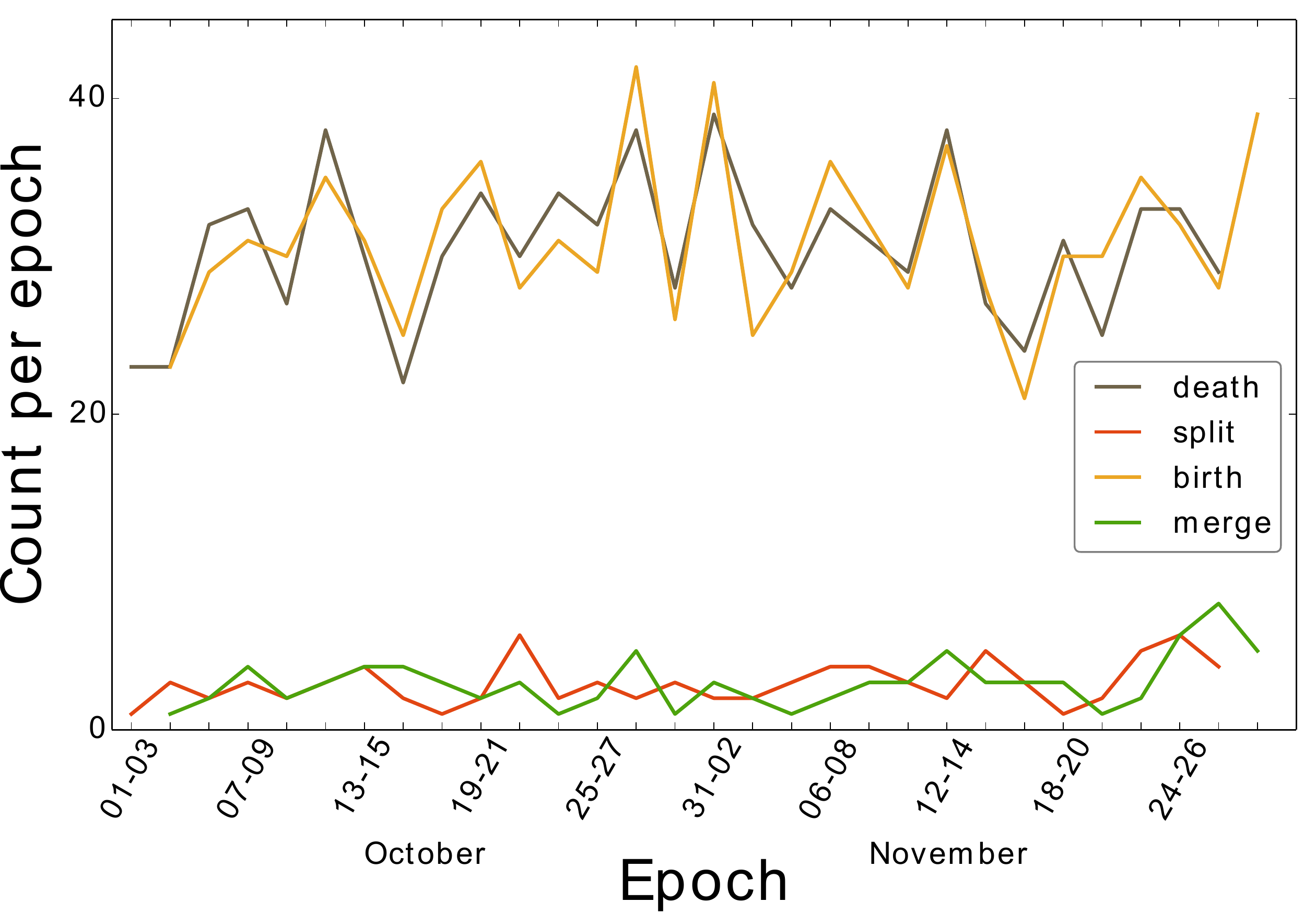}
  \vspace{-5pt}
  \caption{Complex topic evolution statistics (births, deaths, merges, and splits) over a two month period (Oct--Nov 2015). It can be seen that our framework is effective in capturing the highly dynamic nature of Twitter information exchange as witnessed by the changes in the number of inferred topics per epoch across time, consistently observed emergence of new topics and disappearance of others, topic splitting and merging (also see Fig.~\ref{f:progress} for specific examples). Observe the high rate of new topic emergence, compared to the rate of topic creation through the merging of old topics, which illustrates the ephemeral nature of most Twitter exchanges~\cite{HuanThorEfth2010}. }
  \label{f:progressStats}
\end{figure}

Having demonstrated that our method extracts meaningful topics, we next sought to investigate what type of information was obtained by tracking changes in the topic structure over epochs. Here too we readily observed that the proposed method was powerful in extracting interesting and insightful knowledge. Illustrative examples are pictured in Fig.~\ref{f:progress}. The diagrams in Fig.~\ref{f:progress_a} and~\ref{f:progress_b} show examples of ``linear'' topic evolutions i.e.\ changes in the nature of conversation pertaining to a single, continuously developing topic. For example, the evolution in Fig.~\ref{f:progress_a} concerns the bestselling book ``The Reason I Jump: The Inner Voice of a Thirteen-Year-Old Boy with Autism'' written by the autistic Japanese author Naoki Higashida, and which amongst other things discusses the highly emotive issues of language and intelligence~\cite{Higa2013}. On the other hand, Fig.~\ref{f:progress_c} shows multiple topics merging to form new topics. For example, the topic on behavioural and social aspects of schooling can be seen to merge with the topic concerning the already mentioned myth about a link between vaccination and autism, to produce a topic which encompasses a combination of these issues. Lastly, the ability of our framework to capture and track highly dynamic changes in the topic structure, which is crucial in the analysis of a communication medium such as Twitter, is corroborated quantitatively by the plot in Fig.~\ref{f:progressStats}. The plot shows changes in the number of topics per epoch, as well as the corresponding rates of new topic emergence (``birth''), topic disappearance (''death''), merging, and splitting. It can be seen that our framework is effective in capturing the highly dynamic nature of Twitter information exchange as witnessed by the changes in the number of inferred topics per epoch across time, consistently observed emergence of new topics and disappearance of others, topic splitting and merging. It is interesting to observe the high rate of new topic emergence, compared to the rate of topic creation through the merging of old topics, which illustrates the ephemeral nature of most Twitter exchanges~\cite{HuanThorEfth2010}.

\section{Summary and conclusions}\label{s:summary}
In this paper we described the first work on the extraction of the topic structure of tweets and the tracking of its evolution over time. Although our specific aim was that of analysing autism-related Twitter content, the developed framework is entirely general in nature and can be applied to any longitudinal corpus of tweets. Our work identified and elucidated the inherent difficulty associated with topic modelling on tweets -- their insufficient information content. To overcome this limitation we proposed an ingenious framework which uses tweets not as endpoint data mining sources but rather as intermediaries used to discover much richer associated content.

The method proposed in this paper opens a rich cornucopia of possible avenues for further research which we intend to pursue. Some of these concern developments of the overall framework. For example in cases when URL targets include videos (say, on YouTube) the use of computer vision or audio processing can add valuable information for even more sophisticated data mining. On the other hand, there are technical aspects of the proposed method which can be improved further. For example we intend to explore the use of ideas from information-theory for fully automatic similarity graph pruning, which would eliminate the need for the free parameter, in the form of the pruning threshold, described in this work. We also intend to explore automatic ways of labelling topics semantically using meaningful sentence fragments by back-analysing probabilistically the collected text data for persistent ngrams across documents with shared topics~\cite{Aran2012d}.

\balance
{
\bibliographystyle{ieee}
\bibliography{../../../my_bibliography_shrt}
}

\balance

\end{document}